\documentclass[doublecol]{epl2} 
\usepackage{amsmath}

\title{Mobility-induced kinetic effects in multicomponent mixtures}

\author{F. C. Thewes\inst{1} \and M. Krüger\inst{1} \and P.Sollich\inst{1,2}}

\institute{                    
  \inst{1} Institut für Theoretische Physik, Georg-August-Universität Göttingen, 37077 Göttingen, Germany\\
  \inst{2} King's College London, Department of Mathematics, Strand, London WC2R 2LS, U.K.\\
}
\abstract{We give an overview exploring the role of kinetics in multicomponent mixtures. Compared to the most commonly studied binary (single species plus solvent) case, multicomponent fluids show a rich interplay between kinetics and thermodynamics due to the possibility of fractionation, interdiffusion of mixture components and collective motion. This leads to a competition between multiple timescales that change depending on the underlying kinetics. At high densities, crowding effects are relevant and non-equilibrium structures can become long-lived. We present the main approaches for the study of kinetic effects in multicomponents mixtures, including the role of crowding, and explore their consequences for equilibrium and non-equilibrium scenarios. We conclude by identifying the main challenges in the field. 
}

\begin{document}

\maketitle

\section{Introduction}
Starting with the famous phenomenological law of diffusion introduced by Fick almost two centuries ago, the theories behind the kinetics of fluids have evolved from modelling ideal gases to much more complicated scenarios, including the still open problem of glassy dynamics~\cite{Janssen2018-gw}. The movement of individual molecules in a dense fluid is usually subject to severe slowing down arising from e.g.\ volume exclusion, causing strong deviations from ideal behavior and leading to highly non-trivial effects observed in e.g.\ hard-spheres at high packing fractions~\cite{Warren1999_hp,Batchelor1983-mg}.

For multicomponent fluids, which have many different species of particles, the underlying physics becomes even more challenging~\cite{Warren1998-km, Sollich2002-ka,Sollich2007-du,De_Castro2017-yu,Wang2017-ql,Jacobs2013-hg,Mao2019-is,Tafa2001-bz}. Due to liquid-liquid phase separation (LLPS), different regions in space can be enriched in selected mixture components and depleted in others. The kinetic constraints from crowding may then slow down the relaxation~\cite{Warren1998-km}, preventing the system from reaching equilibrium as determined by the usual phase coexistence conditions. A typical example from biophysics is the formation of membraneless structures in the cytoplasm~\cite{Sear2005-zz,Lafontaine2021-iv,Lee2019-po,Alberti2017-dq,Choi2020-rz}, a multicomponent fluid, via LLPS. Due to the high densities~\cite{Sear2005-zz}, the cytoplasm behaves as an aging fluid, showing similarities to glassy systems once its metabolism is suppressed~\cite{Oyama2019, Parry2014} .

Understanding the role of kinetics in multicomponent dense fluids and its interplay with thermodynamics is therefore a rich problem. We will outline here the resulting challenges together with the main theoretical approaches used to tackle them. Our aim is to focus on the interesting phenomena arising specifically from the multicomponent nature of a fluid.

\section{Historical overview: Fick vs.\ Maxwell-Stefan diffusion}
\label{sec1}
We are interested in the \emph{kinetic} properties of multicomponent fluids, so the appropriate ensemble is the canonical (rather than e.g.\ grandcanonical) one. Due to particle number conservation, the time evolution of the system can be written in the form of a continuity equation
\begin{equation}
    \dot{\phi}_i(\boldsymbol{x},t) = -\nabla\cdot \boldsymbol{j}_i
    \label{continEQ}
\end{equation}
where the index $i$ labels the different species in the fluid. The quantity $\phi_i(\boldsymbol{x},t)$ can be interpreted in two ways. On the one hand, from the microscopic perspective, $\phi_i(\boldsymbol{x},t)=\sum_{k\in i}^N\delta(\boldsymbol{x}-\boldsymbol{x}_k)$ is a \emph{density operator} and Eq.~\eqref{continEQ} then has to be read on the same level as the Dean equation~\cite{Dean1996-no}. On the other hand, after coarse graining on a suitable lengthscale (of the order of typical interparticle distances), $\phi_i$ and the current density $\boldsymbol j_i$ become continuous fields. Fick's phenomenological law of diffusion states that the main driving forces for mass transport are density gradients,
\begin{equation}
    \boldsymbol{j}_i = -D_i\nabla\phi_i,
    \label{fick}
\end{equation}
where $D_i$ is the diffusion constant for species $i$. According to this law, mass will flow in the direction of lower densities until a homogeneous profile is obtained in equilibrium. But this relationship is valid only for ideal gases as it accounts exclusively for entropic effects. 

In 1962, Duncan and Toor~\cite{Duncan1962} conducted experiments with ternary ideal mixtures that showed the existence of ``uphill" diffusion, where one of the components would flow in the direction of higher concentration. The experiments were successfully described by the Maxwell-Stefan approach to diffusion~\cite{Krishna1997-xl}. In contrast to Fick's law this uses gradients in the chemical potentials $\mu_i$ as the main driving forces, which have to be balanced by the internal friction between the components. The resulting relation can be stated as
\begin{equation}
    \nabla\mu_i 
    = -\sum_{j=0}^{M} \frac{\phi_j}{\phi_{\textrm{tot}}}\gamma_{ij}(\boldsymbol{v}_i-\boldsymbol{v}_j)
    \label{max-stef}
\end{equation}
On the r.h.s.\ the $\boldsymbol{v}_i=\boldsymbol{j}_i/\phi_i$ are the velocities of the different particle species and $\gamma_{ij}$ is a friction coefficient for relative motion of species $i$ and $j$ (historically often written as the inverse of a Maxwell diffusivity $\tilde{D}_{ij}$), with $\gamma_{ij}(\boldsymbol{v}_i-\boldsymbol{v}_j)$ the corresponding friction force. These forces are then averaged across all species weighted with the appropriate number (or molar) fractions $c_j=\phi_j/\phi_{\textrm{tot}}$, where $\phi_{\textrm{tot}}=\sum \phi_i$ is the total density. The solvent, denoted as species $j=0$, is included on the r.h.s.\ because motion relative to it also generates friction. (We will specify inclusion of the solvent in sums explicitly; otherwise our summation ranges are from $1$ to $M$.)
The set of linear equations~\eqref{max-stef} together with the volume conservation constraint $\sum_{j=0}^M \phi_j \boldsymbol{v}_j=0$ can be inverted to express the velocities $\boldsymbol{v}_i$, and hence the currents $\boldsymbol{j}_i$, linearly in terms of chemical potential differences as
\begin{equation}
    \boldsymbol{j}_i 
    = -\sum_j L_{ij}\nabla\mu_j
    \label{diffusion}
\end{equation}
with all kinetic effects now captured in the (generally density-dependent) mobility matrix $L_{ij}$. When -- as in e.g.\ dilute systems -- the chemical potentials $\mu_i$ are functions of the {\em local} densities $\phi_j$ so that the gradients of these two sets of quantities are linearly related, one retrieves a generalized form of Fick's law~(\ref{fick})
\begin{equation}
    \boldsymbol{j}_i = 
    -\sum_{j,k} L_{ij}\Omega_{jk}\nabla\phi_k 
    \label{diffusion_fick}
\end{equation}
that decomposes the diffusion matrix into the kinetic factor $L_{ij}$ and a thermodynamic factor $\Omega_{ij}$. Because of its more general applicability, the Maxwell-Stefan description has largely superseded the Fickian one in the literature and we therefore follow it below.

\section{Equilibrium Properties}
The problem of phase coexistence in multicomponent mixtures has received significant attention in recent years~\cite{Sollich2002-ka,Mao2019-is,Jacobs2013-hg,Thewes2023}. In general, equality of the chemical potential $\mu_i^\alpha=\mu_i$ of each individual species $i$ among the different phases $\alpha$, as well as equality of osmotic pressure $\Pi^\alpha=\Pi$ leads to the common tangent plane construction for the free energy density
\begin{equation}
    V(\boldsymbol\phi) \equiv f_{\textrm{bulk}}(\boldsymbol{\phi})- \sum_i \mu_i\phi_i + \Pi = 0,
    \label{commonTangent}
\end{equation}
where -- as we are considering bulk properties for now -- $f_{\textrm{bulk}}$ does not contain interfacial terms and we use the shorthand $\boldsymbol\phi = \left (\phi_1, \phi_2, \dots, \phi_M \right )^{\rm T}$. Eq.~\eqref{commonTangent} implies that $P$ different phases will coexist at pressure $\Pi$ and chemical potentials $\mu_i$ if $P$ points on the free energy surface $f_{\textrm{bulk}}$ lie on a common (tangent) plane; the effective potential $V(\boldsymbol\phi)$ measures the distance from this plane~\cite{Sollich2002-ka, Sollich2007-du}. Solving this common tangent plane condition for multicomponent mixtures is a difficult task in general and alternative approaches have been developed, including the moment method~\cite{Sollich2002-ka,Warren1998-km}, grand-canonical simulations~\cite{Jacobs2013-hg} and gradient descent on the total free energy~\cite{Zwicker2022}.

A common choice for the total free energy $F[\boldsymbol\phi]=\int d\boldsymbol x\, f(\boldsymbol\phi, \nabla\boldsymbol\phi; \boldsymbol x) $ for multicomponent mixtures is the so-called Flory or regular solution theory form
\begin{align}
\begin{split}
    f &= T\sum_{i=0}^M \phi_i\log\phi_i + \frac{1}{2}\left [ \boldsymbol\phi^{\rm T}\boldsymbol\epsilon\boldsymbol\phi + \nabla\boldsymbol\phi^{\rm T}\boldsymbol K\nabla\boldsymbol\phi \right ]\\
     &= f_{\textrm{bulk}}(\boldsymbol{\phi}) + \frac{1}{2}\nabla\boldsymbol\phi^{\rm T}\boldsymbol K\nabla\boldsymbol\phi 
    \label{floryFreeEnergy}
    \end{split}
\end{align}
where $\phi_0 = 1-\phi_{\textrm{tot}}$ (in appropriate units) is the solvent density and the elements $\epsilon_{ij}$ and $K_{ij}$ of the matrices $\bm{\epsilon}$ and $\bm{K}$ give, respectively, the pairwise interaction strength between species $i$ and $j$ and the free energy cost of an interface between them. Importantly, the solvent is assumed to be non-interacting, contributing only an entropic term arising from volume exclusion.
 
\section{Kinetics of multicomponent mixtures}
As is clear from the Maxwell-Stefan approach to diffusion, an important question is the interplay of the complex thermodynamic features of multicomponent mixtures with the underlying kinetics, which controls the non-equilibrium properties of the mixture including its approach towards equilibrium. From now on, we focus on this aspect.

\subsection{Collective motion versus interdiffusion}

A key feature of mixture kinetics is the existence of distinct types of motion: the overall {\em particle density} can be equilibrated by {\em collective motion}, where all particle species have similar velocities and hence low friction forces between them. (This collective motion is relative to the solvent that is not written explicitly in $\boldsymbol\phi$; it is not a barycentric flow.) Relaxing the {\em mixture composition}, on the other hand, requires {\em interdiffusion}, i.e.\ motion of different species in opposite directions. This is expected to be much slower at high densities due to crowding effects. Warren proposed that the phase ordering kinetics of mixtures can then be separated into early and late times: initially, the mixture behaves as if its composition were fixed everywhere and separates into phases differing only in density, according to the appropriate ``quenched'' phase diagram. At later times also compositions then equilibrate in line with the conventional ``annealed'' phase diagram~\cite{Warren1998-km,De_Castro2017-yu}. This scenario can be expressed equivalently in terms of moments of the density distribution $\phi_i$: for the free energy~\eqref{floryFreeEnergy} with $\epsilon_{ij}=-\sigma_i\sigma_j$~\cite{De_Castro2017-yu} the total density $\phi_{\textrm{tot}}=\sum_i\phi_i$ relaxes quickly while the composition moments $\sum_i\phi_i\sigma_i^n$ with $n\geq 1$ (more precisely their ratios to $\phi_{\textrm{tot}}$) are slow. Ref.~\cite{De_Castro2017-yu} provided strong evidence for Warren's scenario by showing that the growth rate during the spinodal regime for a polydisperse lattice gas follows the quenched spinodal at high densities, while low density systems follow the annealed one. 

\subsection{Model B dynamics}
In order to capture kinetic effects such as the Warren scenario, we work with a generalization of Eq.~\eqref{diffusion} that includes memory effects~\cite{Rottler2020,Wang2019,Akaberian2023-pj}
\begin{eqnarray}   \dot{\phi}_i(\boldsymbol{x},t) 
&\!\!\!\!\!=\!\!\!\!\!& \sum_j\nabla\cdot \left [ \int d\boldsymbol{x}'\!\!\int_0^t \!dt'  L_{ij} \left (\boldsymbol{x}-\boldsymbol{x}', t-t' \right )  \nabla'\frac{\delta F}{\delta \phi_j} \right ]
\nonumber\\
    &&{}+ \sqrt{2 T}\nabla\cdot {\boldsymbol{\eta}_i}(\boldsymbol{x},t)
    \label{modelB}
\end{eqnarray}
Here the fluctuation-dissipation theorem requires the noise correlations to be $\langle \boldsymbol{\eta}_{i}(\boldsymbol{x},t)
\boldsymbol{\eta}^{\rm T}_{j}(\boldsymbol{x}',t')\rangle = {L}_{ij}(\boldsymbol{x}-\boldsymbol{x}',t-t')\,\boldsymbol{1}$ with $\boldsymbol{1}$ the $d$-dimensional identity matrix. As in Eq.~\eqref{diffusion} the kinetics of the mixture are encoded in the mobility matrix (kernel) $\boldsymbol L$ while the thermodynamics appear as the driving force, namely the gradient of the chemical potential. The specific form of $\boldsymbol L$, as demonstrated in the experiments of Duncan and Toor, is relevant to the non-equilibrium dynamics of the mixture as it can give rise to multiple timescales~\cite{Rottler2020,Akaberian2023-pj}, uphill diffusion~\cite{Casini2023-hv} and indicate the presence of a transition to glassy dynamics. 

\subsection{Mobility matrix}

To obtain a formal expression for the mobility matrix, one can expand the free energy to quadratic order in 
deviations $\delta\phi_i=\phi_i-\bar\phi_i$ from a spatially homogeneous reference state as
\begin{equation}
    F=\frac{1}{2} \sum_{ij}\int d\boldsymbol x \int d\boldsymbol x'\, \delta\phi_i(\boldsymbol x)\Omega_{ij}(\boldsymbol{x}-\boldsymbol{x}')\delta\phi_j(\boldsymbol x')
\label{hamiltonian}
\end{equation}
This gives for (the Fourier transform of) the chemical potentials $\mu_i=\delta F/\delta \phi_i$ the relation $\boldsymbol\mu(\boldsymbol q)=\boldsymbol\Omega(\boldsymbol q)\boldsymbol\phi(\boldsymbol q)$. The matrix $\boldsymbol\Omega(\boldsymbol q)$ linking $\boldsymbol\phi$ and $\boldsymbol\mu$ can now be expressed in terms of structure factors: 
defining the time-dependent structure factor matrix 
$\boldsymbol S(\boldsymbol q,t)=N^{-1}\langle \boldsymbol{\phi}(\boldsymbol{q},t)\boldsymbol{\phi}^{\rm T}(-\boldsymbol{q},0)\rangle$, one has the relation $\boldsymbol\Omega(\boldsymbol q)=(T/\bar\phi_{\rm tot}) \boldsymbol{S}_0^{-1}(\boldsymbol q)$
where $S_0(\boldsymbol q)\equiv\boldsymbol S(\boldsymbol q,t=0)$ is the equal-time structure factor matrix. If one now further assumes $\boldsymbol L$ to be independent of $\boldsymbol \phi$, then one can show~\cite{Akaberian2023-pj} from Eq.~\eqref{modelB}, after taking Fourier and Laplace transforms in space and time, respectively, that
\begin{equation}
    \frac{q^{2}T}{\bar\phi_{\rm tot}}\boldsymbol L(\boldsymbol q,z) =  \boldsymbol S_0 (\boldsymbol q) \boldsymbol S^{-1}(\boldsymbol q,z)\boldsymbol S_0(\boldsymbol q) - z\boldsymbol S_0(\boldsymbol q),
    \label{mobilityGeneral}
\end{equation}
where the $\bar\phi_{\rm tot}$ is a global quantity. For a given mixture one can thus, in principle, measure the required (equal-time and time-dependent) structure factors to determine the underlying mobility matrix. But these are $M\times M$ matrices so such measurements are a difficult task for multi-component mixtures, with additional challenges arising from having to invert such large matrices for all $\boldsymbol q$ and $z$, compounded by the numerical inverse Fourier and Laplace transforms if one wants to obtain the mobility in the time and real space domains. Considerable simplifications can be achieved, however, if one assumes that mobility and thermodynamics are sufficiently decoupled for all particle species to be kinetically equivalent. This approximation, called the painted particle model in Ref.~\cite{Akaberian2023-pj}, yields for the mobility in the limit $z\to 0$
\begin{equation}
    L_{ij}(\boldsymbol q,z\to 0)=
    \frac{\phi_{\textrm{tot}}}{q^2T S^s(\boldsymbol q, z\to 0)}\left [c_i\delta_{ij}  - \left (1-r\right)c_ic_j\right].
    \label{mobilityPainted}
\end{equation}
Similar to $\phi_{\textrm{tot}}$, the $c_{i}\equiv \bar{\phi}_{i}/\phi_{\textrm{tot}}$ are global quantities. Apart from the prefactor, the above expression folds all system-specific information into the parameter $r=S^2_0(\boldsymbol q)S^s(\boldsymbol q, z\to 0)/S(\boldsymbol q, z\to 0)>0$ where $S^s$ and $S$ are, respectively, the self and coherent (scalar) structure factors of the underlying painted particle mixture. For $r>1$ the resulting mobility matrix favors collective motion of particles, as the ratio between diagonal and off-diagonal mobilities is positive. For $r<1$ this ratio is negative and the mobility favors interdiffusion of species (which can be seen as the generalization of the term uphill diffusion used in the Maxwell-Stefan description). The ideal gas limit of Fickian diffusion is recovered for $r=1$, when $\boldsymbol L$ becomes purely diagonal. 

\subsection{Polymers}
The mobility matrix is also important in the mean-field description of polymer mixture kinetics~\cite{Rottler2020,Wang2019,Pagonabarraga2003,Muller2021-xy,Mller2022,Li2021}. As one important example, Ref.~\cite{Pagonabarraga2003} derives the mobility of a polydisperse polymer mixture by minimizing the total dissipation rate with respect to the fluxes of each species, which accounts for the minimization of the free energy~\cite{Weber2019-ep} and the friction between different polymer chains, similarly to the Maxwell-Stefan approach~\eqref{max-stef}. For the polymeric case, the friction was shown to be quadratic in the velocities of the different species and the maximization of the entropy production rate gives the Rayleighian condition 
\begin{equation}    
    \gamma_i(\boldsymbol v_i - \boldsymbol v_t) + \nabla\mu_i\rho_i - N_i\rho_i\sum_{j=0}^M\rho_j\nabla\mu_j = 0
    \label{minEntropy}
\end{equation}
where $\gamma_i$ and $N_i$ are the friction coefficient and the chain length of species $i$, and $\rho_i$ its chain number density. The velocity $\boldsymbol v_t$ is common to all species and can be determined from the incompressibility condition $\sum_{i=0}N_i\boldsymbol j_i=0$. The assumption~\eqref{minEntropy}, together with a microscopic model for the friction $\gamma_i\propto N_i^2\phi_i$, allows one to read off directly the mobility matrix, which has the form
\begin{equation}
    L_{ij}\propto \frac{1}{N_i^2}\phi_i - \left [ \left (\frac{1}{N_i} + \frac{1}{N_j} \right )-\sum_{k=0}\phi_k\right ]\phi_i\phi_j
    \label{mobilityPagon} 
\end{equation}
where the $\phi_i=N_i\rho_i$ are now the chain number densities. An important distinction in this approach is the treatment of the solvent as a regular species, making $\boldsymbol L$ an $(M+1)\times (M+1)$ matrix. A direct comparison to~\eqref{mobilityPainted} is then straightforward only if the limit $\phi_{\rm tot}\to 1$ is taken there.
 
In the scenario of length monodispersity, i.e.\ $N_i=N$, the above mobility matrix was shown in Ref.~\cite{Pagonabarraga2003} to play no role in the kinetics other than to define an overall timescale. This is easily seen by noticing that for $N_i=N$ 
$\boldsymbol L$ has a vanishing eigenvalue with a uniform eigenvector, meaning that any changes in the total density are forbidden by the mobility and only interdiffusion is possible. This happens in~\eqref{mobilityPainted} if $r=0$, which is exactly the case of an incompressible fluid for which $S_0(\boldsymbol q)\to 0$. As soon as the solvent is treated as a passive species, meaning it contributes only entropically to the free energy, all moments relax with different rates and both interdiffusion and collective motion become possible. In the lattice gas section below we will derive the mobility from a microscopic model and show explicitly how these results are recovered.

A more recent approach~\cite{Muller2021-xy,Rottler2020}, similar to the inversion~\eqref{mobilityGeneral}, was used to derive the mobility of polymeric mixtures in Ref.~\cite{Wang2019,Mller2022}. The authors start with the dynamical random phase approximation~\cite{Akcasu1986,Brochard1983} (D-RPA), which gives a linear response relationship between the density fields of each type of monomer and a set of time-dependent external potentials $\boldsymbol V=\{V_i\}$:
\begin{equation}
    \boldsymbol\phi(\boldsymbol q, t)  = \phi_{\rm tot}\int_{-\infty}^tdt'\,\frac{d\boldsymbol S(\boldsymbol q,t-t')}{dt}\boldsymbol V(\boldsymbol q,t').
    \label{RPA-LRT}
\end{equation}
Assuming the potentials are constant up to $t'=0$ and then switched off,
the above expression integrates to $\boldsymbol \phi(\boldsymbol q,t) = -\phi_{\rm tot} \boldsymbol S(\boldsymbol q,t)\boldsymbol V(\boldsymbol q)$ and differentiating w.r.t.\ time yields the Onsager relation $\dot{\boldsymbol \phi}(\boldsymbol q,t) = \dot {\boldsymbol S}(\boldsymbol q,t)\boldsymbol S^{-1}(\boldsymbol q,t)\boldsymbol \phi(\boldsymbol q,t)$. The Laplace transform of this relation is then consistent with the Fourier-Laplace transform of~\eqref{modelB} only if $\boldsymbol L$ is given by~\eqref{mobilityGeneral}.

While yielding effectively the same expression for the mobility, the interpretation of Eq.~\eqref{mobilityGeneral} as derived in~\cite{Akaberian2023-pj} is quite distinct from the one in~\cite{Mller2022}. The former assumes a painted particle scenario and uses as input the \emph{total} dynamical structure factor of the \emph{interacting} system. Any memory thus emerges from \emph{inter}molecular interactions. The latter~\cite{Mller2022} starts with D-RPA, which takes the ideal gas as a reference frame where total and single molecule structure factors agree. The input to the mobility is this \emph{single molecule} dynamical structure factor and memory emerges from \emph{intra}molecular interactions. The first is more suitable for strongly interacting colloidal systems while the second best describes weakly interacting polymer mixtures. The physical assumption underlying both approaches is that of a near-equilibrium situation, which implies a linear relation between driving forces and responses.

In order to derive more concrete results for the mobility one needs to provide the microscopic information contained in the dynamical structure factors $\boldsymbol S(\boldsymbol q, z)$. In Refs.~\cite{Mller2022,Muller2021-xy,Wang2019,Li2021} these were assumed to be given by the Rouse model. The resulting mobility was shown to change the spinodal dynamics after a quench of a polymeric mixture: for times short compared to the single chain relaxation time, an initial density modulation will decay faster than the usual diffusive relation $
\phi(\boldsymbol q,t)\sim e^{-D\boldsymbol q^2t}$. This faster relaxation is directly related to the presence of memory in the mobility kernel, an effect that was not considered in~\cite{Pagonabarraga2003}.

\subsection{Relation to Dynamical Density Functional Theory}
The agreement of the expressions for the mobility in the previous sections indicates a more subtle connection. In the famous Dean equation~\cite{Dean1996-no}, the time evolution of the density \textit{operators} $\phi_i(\boldsymbol x)$ is obtained in terms of the pairwise potential between individual particles and a diagonal mobility $L_{ij}=\phi_i\delta_{ij}$ depending only on the local densities. This equation, although exact at the level of density operators, is challenging to apply in practice as e.g.\ it fails to recover the correct equilibrium structure factors if applied naively to smooth density fields. This issue is resolved in dynamical density functional theory~\cite{Fraaije1997,Marconi2000-qo,Te_Vrugt2020-vu} by introducing a free energy functional $\mathcal{F}[\boldsymbol\rho]$ dependent on {\em average} densities $\rho_i(\boldsymbol x)=\langle \phi_i(\boldsymbol x)\rangle$. By construction such a theory will yield exact results for the {\em static} fluctuations~\cite{Kruger2017-qn,Akaberian2023-pj} of the $\phi_i$ (to quadratic order), if the appropriate free energy~\eqref{hamiltonian} is used with~\cite{Kruger2017-qn}
\begin{eqnarray}
\beta \Omega_{ij}(\boldsymbol x - \boldsymbol x') &=& 
\beta\left.\frac{\delta^2 {\mathcal{F}}}{\delta \rho_i(\boldsymbol x)\delta \rho_j(\boldsymbol y)}\right|_{{\boldsymbol\rho}(\boldsymbol x)=\bar{\boldsymbol\phi}}\\
&=&
\frac{\delta_{ij}}{\bar\phi_i}\delta(\boldsymbol x-\boldsymbol x') - c_{ij}^{(2)}(\boldsymbol x-\boldsymbol x')
\label{Omega_c}
\end{eqnarray}
where $\boldsymbol c^{(2)}$ is the direct pair correlation function. This expression makes explicit the separation of $\boldsymbol\Omega$ into a local ideal gas contribution in the first term and a nonlocal contribution from $\boldsymbol c^{(2)}$ in the second~\cite{Kruger2017-qn}. The key  additional benefit of using the mobility \eqref{mobilityGeneral} is then that also the time-dependent quadratic fluctuations of the $\delta\phi_i$ are predicted exactly.

By linear response, the same then also applies to the relaxation dynamics of the {\em average} density $\rho_i$ of DDFT: the following adjusted equation for DDFT,
\begin{equation}
\dot{\boldsymbol\rho}(\boldsymbol{x},t) = \nabla\cdot \left [ \int d\boldsymbol{x}'\int_0^t dt'  \boldsymbol L \left (\boldsymbol{x}-\boldsymbol{x}', t-t' \right )  \nabla'\frac{\delta {\mathcal{F}}}{\delta \boldsymbol\rho(\boldsymbol x',t')} \right ]
    \label{DDFT}
\end{equation}
yields the exact relaxation dynamics close to equilibrium, i.e.\ to leading order in $\rho_i(\boldsymbol x,t) -\bar{\phi}_i$.

In summary, if the time evolution of the structure factors of a given system is known, the ``correct'' mobility can be determined from~\eqref{mobilityGeneral}. This mobility, together with the appropriate free energy~(\ref{hamiltonian},\ref{Omega_c}), generates the exact (second order) fluctuation statistics of the $\phi_i$, both statically and dynamically, when used in the generalized model B equation~\eqref{modelB}. The corresponding generalized DDFT equation (\ref{DDFT}) similarly predicts correctly the time-dependent linear responses. Additionally, non-trivial features of the mobility as discussed above allow for the possibility to account for the Warren scenario, as well as arrested dynamics or the Mode Coupling glass transition~\cite{Janssen2018-gw}. Exploring the consequences of the full mobility expressions in the predictions of dynamical density functional theory thus offers a promising direction for future research.

\subsection{Microscopic approach: the mobility of a lattice gas}
In the previous sections we have reviewed general approaches for obtaining the mobility matrix from macroscopic behavior captured in the dynamical structure factors. Here we use a different route and start from a microscopic model consisting of a multicomponent lattice gas with Hamiltonian
\begin{equation}
    H = \sum_{\boldsymbol x\sim \boldsymbol x'} \sum_{i,j=1}
n_i(\boldsymbol x) n_j(\boldsymbol x') \epsilon_{ij}
\end{equation}
where $n_i(\boldsymbol x)$ is the occupancy number of site $\boldsymbol x$ by species $i$, i.e.\ equal to one if the site is occupied by a particle of species $i$ and zero otherwise. The sum over $\boldsymbol x\sim \boldsymbol x'$ runs over nearest neighbor pairs. Finally, the energy cost of a bond between species $i$ and $j$ is given by $\epsilon_{ij}$. We assume additional hard core interactions at short distances such that each lattice site can be occupied by at most one particle.

In the supplementary material~\cite{suplMat} we derive the hydrodynamic limit of this model using a path integral approach~\cite{Lazarescu2019} and show that the resulting equations of motion have the form~\eqref{modelB}, such that the mobilities $L_{ij}$ can be read off directly
\begin{equation}
    L_{ij}(\boldsymbol x) =  \frac{\phi_{\textrm{tot}} \sum_{k=0}^M w^ {ik}\phi_k }{T}\left [ c_i\delta_{ij} - \frac{\phi_{\textrm{tot}}w^{ij}c_i c_j}{\sum_{k=0}^M w^ {ik}\phi_k} \right ]
    \label{mobilityLatticeGas},
\end{equation}
where $w^{ij}$ is the rate at which two neighboring particles of type $i$ and $j$ swap positions. This expression is the multicomponent generalization of the binary case studied in~\cite{Kehr1989}.

The mobilities~\eqref{mobilityLatticeGas} shed further light on the expressions for $\boldsymbol L$ in Eq.~\eqref{mobilityPainted} and~\eqref{mobilityPagon}. Since we assumed Markovian dynamics, the resulting mobility is local in time, while locality in space follows from the hydrodynamic scaling. The quantities $\phi_i$, $\phi_{\rm tot}$ and $c_i$ on the r.h.s.\ are therefore now {\textit local}, i.e.\ evaluated at position $\boldsymbol x$, in contrast to  Eq.~\eqref{mobilityPainted} where the  \textit{global} quantities enter. 

By setting $w^{ij}=w_s$ $\forall i,j\neq 0$ and $w_{i0}=w_0$ $\forall i$, 
a mobility matrix of painted particle form~\eqref{mobilityPainted} is recovered from Eq.~\eqref{mobilityPainted}, consistent with the idea that in the painted particle picture all particle species (but not vacancies) are kinetically equivalent.
The (local) parameter $r$ is then given by $r=w_0\phi_0/[w_s\phi_{\textrm{tot}} + w_0\phi_0]$. This can be read as the ratio between the scales for the current from collective motion,  $w_0\phi_0 \phi_{\rm tot}=w_0\phi_{\rm tot}(1-\phi_{\rm tot})$, and for the total current $w_0\phi_0\phi_{\rm tot} + w_s\phi_{\textrm{tot}}^2$ that also contains the interdiffusion contribution $w_s\phi_{\textrm{tot}}^2$. Due to the hydrodynamic limit, $r\leq 1$ and collective motion ($r\to 1$) dominates for $w_s\to 0$, i.e.\ when particle swaps become suppressed. Nonetheless, for any non-vanishing $w_s$ one has that $r\to 0$ for $\phi_{\rm tot}\to 1$, which is consistent with our results above for the length-monodisperse polymeric mobility and the painted particle mobility in the incompressible limit.

\section{Coarsening: the role of kinetics}
Up to now we have looked at general forms of the mobility. In the following sections we focus on the effects of these kinetic properties on specific physical processes. 

We start by investigating the late stages of phase separation, where an important question regarding kinetic effects arises, namely their role in the coarsening process~\cite{Puri1997,Bray1995,Emmott1999,Svoboda2010-fv,Philippe2013-ji,Lee2021,Wang2017-ql}. Once a system is brought into a coexistence region of its phase diagram and droplets of a new phase have been formed via either nucleation or a spinodal instability, coarsening or Ostwald ripening~\cite{Kuehmann1996-dg} will set in, with bigger droplets growing by diffusive transport in favor of smaller ones. Here, we highlight the importance of the kinetic aspects in this process for a multi-component mixture.

We show in the supplementary material~\cite{suplMat} that combining the generalized Gibbs-Thomson boundary condition and the solution to the Laplace equation for a spherical droplet of curvature $G=2/R$ one finds for the time evolution of the radius $R$ of a droplet of phase $\alpha$ immersed in a sea of phase $\gamma$ (see~\cite{Svoboda2010-fv,Philippe2013-ji} for alternative derivations)
\begin{equation}
    \frac{dR}{dt} = \frac{2\sigma_{\alpha\gamma}}{R\Delta\boldsymbol\phi^{\rm T} (\boldsymbol L^{\gamma})^{-1}\Delta\boldsymbol\phi}\left (\frac{1}{R_c} - \frac{1}{R}\right)
    \label{drdt}
\end{equation}
with $R_c=2\sigma_{\alpha\gamma}/\Delta \Pi$ the critical radius, while $\sigma_{\alpha\gamma}$ is the surface tension between phases $\alpha$ and $\gamma$. This result retrieves the known  $R\sim t^{1/3}$~\cite{Bray1994-yi} scaling but with a prefactor depending on the compositions of the droplet and the surrounding phase, as well as the mobility in the latter. 

To get some quantitative insight into the effects of this prefactor, consider a painted particle mobility, i.e.\ $\boldsymbol L^\gamma$ of the form~\eqref{mobilityPainted}, $L_{ij} \propto c_i^\gamma \delta_{ij} - (1-r)c_i^\gamma c_j^\gamma$, in the surrounding phase $\gamma$. The matrix inverse can then be performed to give
\begin{equation}
    \Delta \boldsymbol\phi^{\rm T} \boldsymbol L^{-1}\Delta\boldsymbol \phi \propto \sum_i \frac{(\Delta \phi_i)^2}{c_i^\gamma} + \frac{1-r}{r}(\Delta\phi_{\textrm{tot}})^2 
    \label{drdtRate}
\end{equation}
If phase $\gamma$ has a reasonably uniform composition so that $c_i^\gamma=O(1/M)$, the scaling of the first term depends on the $\Delta\phi_i$: if there is strong fractionation~\cite{Thewes2023} so that a few $\Delta\phi_i$ are $O(1)$, the term is $O(M)$ and hence the coarsening rate $O(M^{-1})$. For weak fractionation, on the other hand ($\Delta\phi_i=O(1/M)$), the prefactor~\eqref{drdtRate} is $O(1)$ and the coarsening rate larger by a factor of $O(M)$. Consistent with the Warren scenario, this result suggests that coarsening will proceed first via condensation of a weakly fractionated phase, with growth of strongly fractionated phases occurring only at late times. Conversely, a nucleated domain will shrink much more slowly if it is enriched in only a few mixture components; this makes sense as interdiffusion of particles is required to equilibrate the surrounding composition, while relaxing the total density is much faster. In the limit $r\to 0$ of pure interdiffusion, which occurs for dense systems (see discussion of lattice gas mobility above), the second term in \eqref{drdtRate} always dominates and the Warren scenario disappears.

A second effect coming directly from the mobility was first demonstrated by Bray~\cite{Bray1995} for the case of a single component system. Consider a droplet of phase $\alpha$ immersed in a sea of phase $\gamma$ in which the mobility $\boldsymbol L^{\gamma}$ vanishes. In the case of a single component system and mobility $\propto |(\phi_{\rm tot}^\gamma-\phi_{\rm tot})(\phi_{\rm tot}-\phi_{\rm tot}^\alpha)|^n$, which vanishes in both bulk phases, Bray showed that coarsening proceeds with an anomalous power-law scaling $R\sim t^{1/(3+n)}$. Numerical simulations of this mobility and extensions to non-symmetric cases~\cite{Zhu1999,Dai2012,Dai2016}, i.e.\ with the mobility vanishing only in one of the bulk phases, showed a range of different exponents for the growth of droplets. For a multicomponent system where the mobility $L_{ij} \propto (\phi_{\textrm{tot}} - \phi_c)^n$ vanishes only at a critical total density $\phi_c$ equal to the density of the densest phase, a similar range of exponents as observed in~\cite{Zhu1999,Dai2012,Dai2016} is expected. If $\phi_c$ is lower than the density of some equilibrium phases, as occurs in glass~\cite{Janssen2018-gw} and jamming~\cite{Ikeda2012,Ikeda2013} transitions or in the Brazil nut effect~\cite{Levin2001-gj}, the situation becomes highly non-trivial and the system will generally become arrested in non-equilibrium states~\cite{Sappelt1997,Puri1997,Dai2012,Dai2016}. Scenarios where the mobility vanishes due to changes in composition instead of total density remain unexplored and are likely to be even more challenging to analyze.

\section{Chemical reactions and kinetics} Extending the dynamics~\eqref{modelB} to include chemical reactions is relatively straightforward within a mean-field approach~\cite{Weber2019-ep,Zwicker2022-qg}. 

To illustrate the importance of the kinetics for chemical reactions, consider a mixture with thermodynamically ideal species where the free energy consists only of the ideal gas entropy contribution so that $\Omega_{ij}(\boldsymbol q)\sim \delta_{ij}/\phi_i$. If the mixture undergoes a thermal quench, non-trivial correlations between species can nonetheless appear if the mobility matrix favors interdiffusion~\cite{Akaberian2023-pj}. These correlations can then facilitate reactions that would otherwise take much longer to occur, by concentrating reactants near each other. For such an effect to be relevant, the typical diffusion timescales, which are determined by the eigenvalues of $\boldsymbol L\boldsymbol \Omega$, must be longer than the typical reaction times such that multiple reactions can occur during the lifetime of the kinetic correlations.

A particular case of the mobility-induced (sic!) phase separation discussed above was studied in~\cite{Cotton2022} for a ternary mixture composed of a catalyst ($e$) and two reacting species ($s$ and $p$). The mixture was coupled to reservoirs of fuel and waste held at constant (but distinct) chemical potentials and the mobilities of the three independent species were chosen phenomenologically to be $L_{ij} = \frac{1}{T}\left ( \sum_k D_{ik}\phi_k\phi_i\delta_{ij}-D_{ij}\phi_i\phi_j\right )$, which is exactly~\eqref{mobilityLatticeGas} if $D_{ij}=w^{ij}$, while the mobilities of fuel and waste are unspecified since they play no role in the equations of motion. The authors showed that macroscopic phase separation occurs for $D_{pe}\neq D_{se}$. This choice, due to the distinct chemical potentials of fuel and waste, creates a positive feedback that amplifies any inhomogeneities in the concentration of the catalyst.

In the opposite limit, where diffusion is much faster than reactions, an effective free energy that incorporates creation-annihilation processes can be written by using an analogy with electrostatics, resulting in a long range Coulomb-like interaction that stabilizes the size of phase separated domains~\cite{Glotzer1995}. For such systems, non-local effects in space and time in the mobility become more relevant, pointing towards an interesting direction for future exploration.

\section{Perspectives}
We conclude with a selection of open questions and challenges:
\begin{itemize} \setlength{\itemsep}{0.5pt}
    \item As we have seen, mobilities are generically nonlocal in space and time, and efficient simulation techniques are needed to deal with this.    
    \item The general mobilities we presented are valid for homogeneous mixtures, while the lattice gas approach introduces a hydrodynamic scaling implying a typical coarse-graining length scale. For non-homogeneous mixtures, what is the optimal coarse-graining scale for the density fields such that the general expressions for the mobility are still valid?
    \item What new physics arises from the known mixture free energy functionals of dynamic density functional theory once crowding and dynamical arrest effects are accounted for via the mobility in the generalized model B?
    \item How do memory effects from the mobility affect nucleation and growth dynamics? Once the lengthscale of nonlocality of the mobility becomes of the same order as the critical droplet radius, very interesting new physics should appear.        
    \item Active and non-reciprocal mixtures: these generally have non-equilibrium steady states, leading one to expect that kinetic effects will become even more relevant and potentially key to determining the nature of the steady states. The role of mobility in active~\cite{Cates2023-br,cates2022active} or non-reciprocal systems~\cite{Saha2020-xn} remains largely unexplored.
    \item Studying the influence of mobilities on fluid viscosity~\cite{krugerdean2017b} is an interesting topic for future work.
\end{itemize}

In summary, accounting for non-trivial kinetic effects in multicomponent mixtures leads to fascinating physics that has no direct counterpart in single species systems or, essentially equivalently, incompressible binary fluids. While we have given an overview of the underlying
theory behind such kinetic descriptions, major fruitful challenges
remain both in theoretical as well as numerical approaches.

\section{Acknowledgements} The authors thank Marcus Müller for helpful discussions. This work was supported by the German Research Foundation (DFG) under grant numbers SO 1790/1-1 and KR 3844/5-1.

\bibliographystyle{eplbib}

\end{document}